# Electronic and optical properties of b-AsP quantum dots and quantum rings under the electric and magnetic fields

Chen Li[a], Xuefei Yan[b], Jinluo Cheng[c,d]*, Weiqi Li[a]*

[a] *School of Physics, Harbin Institute of Technology, Harbin 150001, China*
[b] *School of Microelectronics Science and Technology, Sun Yat-Sen University, Zhuhai 519082, People's Republic of China*
[c] *The Guo China-US Photonics Laboratory, Changchun Institute of Optics, Fine Mechanics and Physics, Chinese Academy of Sciences, 3888 Eastern South Lake Road, Changchun, Jilin 130033, China*
[d] *School of Physical Sciences, University of Chinese Academy of Sciences, Beijing 100049, China*

**Abstract:** In this work, we investigate the electronic and optical properties of the b-AsP quantum dots(QDs) and quantum rings(QRs) with different edge types in the presence of an in-plane electric field and a perpendicular magnetic field utilizing the tight-binding method. Our calculations show that the electronic and optical properties of edge states largely depend on the edge types. By adjusting the intensity of the electric field or magnetic field, the probability density can be effectively tuned. In particular, under an in-plane electric field, novel edge states emerge in rhombus QDs and QRs with $\delta$ atoms, and transitions happen between these novel edge states and bulk states in the conduction band region, which is illustrated by distinct and evenly separated peaks in the optical absorption spectra. This work might exploit a potential system to implement quantum state transfer among two-dimensional material quantum dots.

# 1 Introduction

Two-dimensional (2D) arsenic–phosphorus (AsP), as a derivative of black phosphorus (BP), has attracted a lot of attention due to its unique electronic and optical properties. It can be achieved by introducing arsenic into phosphorus via various processes. Similar to phosphorus, arsenic phosphorus has different phases such as $\alpha$-AsP and $\beta$-AsP, some of which proved to be thermally stable[1]. The bandgap of AsP has been found to be experimentally tunable from around 0.15eV to 1.69 eV by component and layer number[2]. Up to now, many electronic and optical devices based on AsP have been reported, for example, $\alpha$-AsP has ultrahigh carrier mobility[3], tunable bandgap[2], component dependent optical response[4], so it has been applied to field effect transistors (FETs)[6], gas sensors[5], photonics devices[7] and mid-infrared photodetectors[8].

Quantum dots (QDs) generally refer to low-dimensional semiconductor nanostructures that are subject to the quantum confinement effect in three spatial directions. Up to now, the electronic and optical properties of quantum dots made from two-dimensional materials have been studied extensively, which are found to be largely dependent on edge configurations[9-13], size[14], and they show various responses to the electric and magnetic field[15-18]. Owing to their sensitivity to the field and incident light, many applications based on quantum dots have been reported such as optical modulators [19], photocatalytic[21], solar cells[20], and single-photon sources[22].

Recently, Li *et al* predicted a puckered and dynamically stable monolayer Janus b-AsP, which is different from the already synthesized and predicted structures of b-AsP in an alloy form[23]. The monolayer Janus b-AsP was found to possess a direct band gap dominated by the P atoms. It exhibits significant in-plane absorption anisotropy due to its puckered lattice structure. A sixteen-band tight-binding (TB) model is also presented for Janus single-layer b-AsP, and based on this TB model, Yan *et al* studied the electronic and optical properties of a square Janus b-AsP quantum dots under a magnetic field[24]. However, some unique properties of b-AsP quantum dots have not been explored yet.

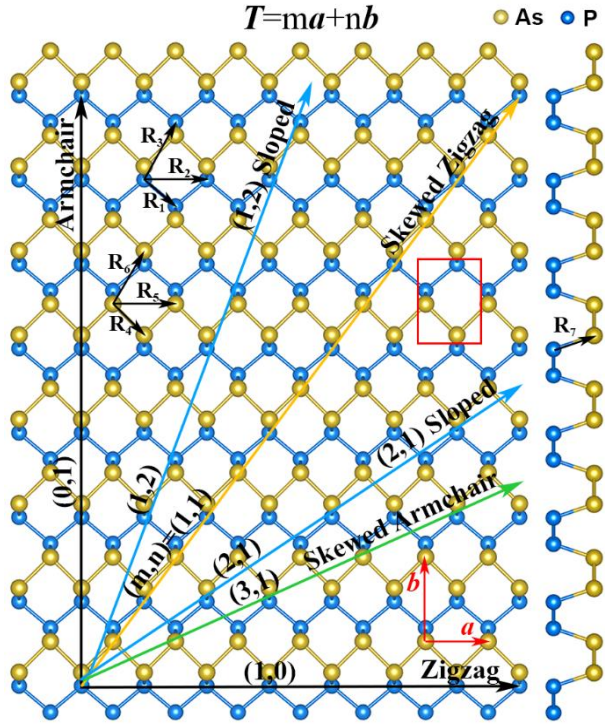

**Fig.1.** Schematic illustration of monolayer b-AsP QD. The red rectangle refers to the unit cell of b-AsP; **a** and **b** are the lattice constants; the chiral vector **T** is illustrated by the thick solid arrow along each 2D crystallographic direction; $R_1$-$R_7$ are the seven hopping parameters.

In this work, we compare the electronic and optical properties of Janus b-AsP quantum dots(QDs) and quantum rings(QRs) with different edge types under the perpendicular magnetic field and the in-plane electric field. The main results obtained in this work are as follows: (i) Edge states appear in the energy gap for QDs with different edge types. The P and As edge atoms can be categorized into $\delta$ and $\gamma$ atoms, and the electronic properties of edge states dominated by different types of atoms vary noticeably. (ii) The energy spectra and the probability densities can be tuned by the in-plane electric field, and novel edge states emerge in rhombus QDs and QRs with $\delta$ atoms in which the probability densities of different energy levels reside on atoms with different potentials. (iii) The probability densities evolve under the increasing perpendicular magnetic field, and the energy levels of ring states oscillate with the magnetic flux. (iv) In the presence of an electric field, transitions happen between novel edge states and bulk states in the conduction band regime, which is illustrated by distinct and evenly separated peaks in the optical absorption spectra. Our work is organized as follows: In section 2, we introduce QDs and QRs with various edge configurations and give detailed descriptions of the construction of the TB Hamiltonian of monolayer b-AsP and illustrate how the magnetic and electric fields are taken into account. In section 3, we compare and discuss the influence of magnetic and electric fields on QDs and QRs with various edge types and explore how the electronic probability densities evolve under these external fields and calculate the optical spectrum of rhombus QDs and QRs in the presence of an in-plane electric field. Finally, we summarize the main conclusions in section 4.

## 2 Model and Theory

Fig.1 illustrates the crystal structure of a monolayer b-AsP and some directions defined by chiral vectors

$$\mathbf{T} = m\mathbf{a} + n\mathbf{b} \tag{1}$$

where **a** and **b** are the primitive vectors with amplitudes |**a**| = 0.347nm and |**b**| = 0.472 nm, respectively. The edge configuration discussed in this work include zigzag (ZZ) for (m,n)=(1,0), armchair (AC) for (0,1), skewed zigzag(SZ) for (1,1), skewed armchair (SA) for (3,1), and sloped edges (SL) for other (m,n) [25]. Along each direction, depending on the number of the dangling bonds of the outmost atoms, the edges can be divided into bearded edges for two dangling bonds and nonbearded edges for one dangling bond [26]. In this work, we consider only more stable QDs

and QRs with nonbearded edges, and name a As(P) atom with As-As(P-P) dangling bond as γ-As (γ-P) while a As(P) atom with As-P dangling bond as δ-As(δ-P).

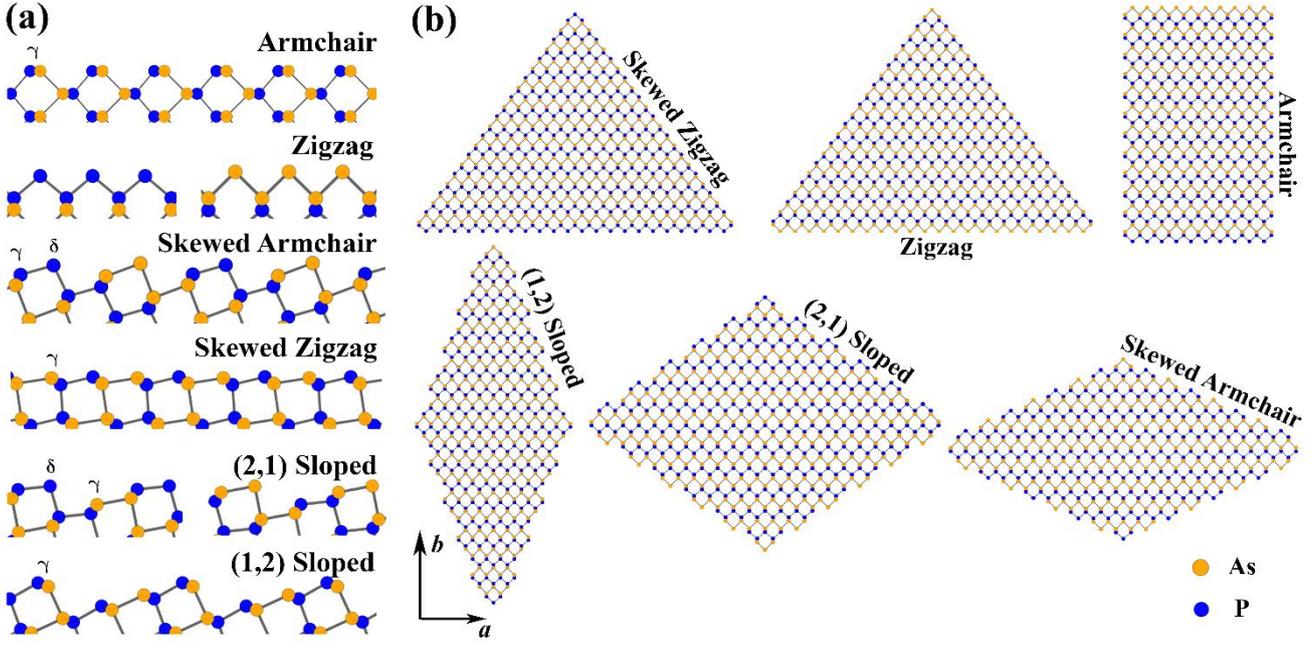

**Fig.2.** (a) An illustration of six typical edges, Armchair(AC), Zigzag(ZZ), Skewed Armchair(SA), Skewed Zigzag(SZ), (2,1) Sloped(SL), (1,2) Sloped(SL). SA and (2,1) SL both have δ and γ atoms while ZZ with only δ atoms and AC, SZ and (1,2) SL with only γ atoms.(b) a collection of QDs with different shapes and edges. They are PTQD/AsTQD with SZ and ZZ edges, SQD with AC and ZZ edges, (1,2) RQD with (1,2) SL edges, (2,1) RQD with (2,1) SL edges and (3,1) RQD with SA edges.

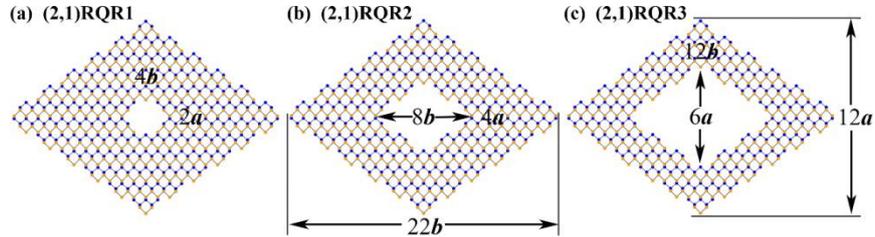

**Fig.3.** A collection of (2,1) RQRs with different widths (the distance between the inner and outer edge).

Table I. Basic quantities of the investigated b-AsP QD.

| Shape | Boundary | $N_{edge}$ | $L_{edge}$/nm | $N_{total}$ | $S/nm^2$ |
|---|---|---|---|---|---|
| PTQD/AsTQD | ZZ | 23 | 7.29 | 573 | 19.82 |
| | SZ | 44 | 12.88 | | |
| SQD | ZZ | 20 | 6.25 | 504 | 18.02 |
| | AC | 48 | 10.38 | | |
| (1,2)RQD | (1,2)SL | 78 | 19.49 | 494 | 17.12 |
| (2,1)RQD | (2,1)SL | 70 | 19.01 | 576 | 21.62 |
| (3,1)RQD | SA | 70 | 19.32 | 478 | 18.43 |

The configurations of the six types of edges being studied in this work are schematically shown in Fig.2(a). All these edge types except zigzag edge have γ-P and γ-As atoms and they appear alternately along the edges. ZZ/(2,1)SL edges can be divided into two categories, δ-P dominated or δ-As dominated, while SA edge both have δ-P and δ-As atoms and they appear alternately on the edge. Fig.2(b) shows a collection of differently shaped flakes whose edges cover the six edge types. They are P/As dominated triangular quantum dot (P/AsTQD), square quantum dot(SQD) and (1,2)/(2,1)/(3,1) rhombus quantum dot [(1,2)/(2,1)/(3,1)RQD]. PTQD and AsTQD both have SZ edges and their zigzag edges are dominated by P and As atoms, respectively. (1,2)RQD is composed of four (1,2)SL edges, while (2,1)RQD and (3,1)RQD are composed of four (2,1)SL and SA edges, respectively. Note that δ-P and δ-As atoms appear apart on two sides of (2,1)RQD while they appear alternately on the four edges around the (3,1)RQD. The basic information including the number of edge atoms $N_{edge}$, the number of total atoms in QDs $N_{total}$, the side length $L_{edge}$, and the surface area $S$ are given in Table I for the QDs. In this work, we also study the electronic and optical properties of (2,1)rhombus quantum rings[(2,1)RQRs], and the basic information of them are shown in Fig.3.

Table II. TB parameters for b-AsP in an orthogonal $sp^3$ model.

| R | $V_{ss\sigma}$ | $V_{sp\sigma}$ | $V_{pp\sigma}$ | $V_{pp\pi}$ |
|---|---|---|---|---|
| $R_1$ | -2.114 | -4.053 | 3.607 | -1.442 |
| $R_2$ | 0.169 | 0.251 | -0.500 | -0.010 |
| $R_3$ | -0.503 | -0.352 | 0.160 | -0.138 |
| $R_4$ | -1.219 | -2.587 | 3.107 | -0.570 |
| $R_5$ | -0.249 | -0.084 | 0.739 | -0.131 |
| $R_6$ | -0.046 | -0.337 | 1.280 | -0.249 |
| $R_7$ | 1.839 | 1.226 | -3.133 | 1.017 |

Table III. The on-site energies for As and P atoms (in eV).

| | $s$ | $p_x$ | $p_y$ | $p_z$ |
|---|---|---|---|---|
| As | −10.446 | −0.756 | −0.346 | −0.760 |
| P | −8.047 | 1.537 | 1.116 | −2.724 |

In this work, We use the Slater-Koster tight-binding (TB) model fitted from the first-principles calculation to describe the low-energy electronic and optical properties of monolayer Janus b-AsP. The basis used in this TB model is $sp_3$ orbitals. In this model, the monoelectronic Hamiltonian and the overlap matrix can be written as

$$H_{mv,m'v'} = \langle \phi_v(r - r_m) | H | \phi_{v'}(r - r_{m'}) \rangle,$$
$$S_{mv,m'v'} = \langle \phi_v(r - r_m) | \phi_{v'}(r - r_{m'}) \rangle, \quad (2)$$

where *m, m'* and *v, v'* run over the lattice sites and atomic orbits, respectively. In the TB model, the interatomic matrix elements between atoms can be calculated using the Slater-Koster method:

$$\langle s|H|s \rangle = v_{ss\sigma},$$
$$\langle s|H|p_i \rangle = n_i v_{sp\sigma}, \quad (3)$$
$$\langle p_i|H|p_j \rangle = (\delta_{ij} - n_i n_j) v_{pp\pi} + n_i n_j v_{pp\sigma},$$

here, $v_{ss\sigma}$, $v_{sp\sigma}$, $v_{pp\sigma}$, $v_{pp\pi}$, $S_{ss\sigma}$, $S_{sp\sigma}$, $S_{pp\sigma}$, $S_{pp\pi}$ are given in Table III, *i* and *j* represent x, y and z. $n_i$ = **R**·**e**$_i$/|**R**| refers to the directional cosines, where ***R*** is the vector along the bond. Moreover, the rule of the angular quantum number[<l|H|l'>=(-1)$^{l+l'}$<l'|H|l>] should be considered to evaluate conjugate transpose matrices. Therefore, the hamiltonian matrix elements can be obtained through equation(3) using the parameters in Table II and Table III.

When a perpendicular magnetic field B is applied to the monolayer b-AsP, the Peierls substitution should be introduced, and the hamiltonian matrix element becomes

$$H_{mv,m'v'} \rightarrow H_{mv,m'v'} \exp\left(i\frac{2\pi e}{h} \int_{\mathbf{r}_i}^{\mathbf{r}_j} \mathbf{A} \cdot d\mathbf{l}\right), \quad (5)$$

where *h* is Planck's constant and A is the vector potential induced by the field B. We use the Landau gauge, with the vector potential A = (0, *Bx*, 0). The magnetic flux threading the b-AsP unit cell is defined as Φ = *Bab* in units of the flux quantum $\Phi_0$ = *h/e*, where ***a*** and ***b*** are the lattice constants.

When an in-plane electric field is applied to the quantum dots/rings, the on-site energy of atoms will change accordingly. If we set the potential of atom i to zero, the on-site energy of other atoms should change from the original ε to ε + *e*E · R$_{ij}$, where R$_{ij}$ is the vector pointing from the zero-potential atom i to atom j.

The electronic density of state (DOS) of the QD system is expressed by the sum of delta function centered on the eigenenergies of the system

$$\text{DOS}(E) = \frac{1}{\Gamma\sqrt{2\pi}} \sum_n e^{\left[-(E-E_n)^2/2\Gamma^2\right]}, \quad (4)$$

where $\Gamma$ is the broadening factor and $E_n$ represents the eigenvalues for the $n$th eigenstate.

The optical absorption coefficient can be calculated through equation (6). Assuming that the light is polarized in the plane $(x, y)$, the dipole matrix element for the transition from the initial state |i> to the final state |j> can be defined as $M_{ij}$ = <j|r|i>. Therefore, the optical spectrum of the b-AsP QD can be calculated by

$$A_{ij}(\hbar\omega) = \sum_{i,j} (E_j - E_i) |\mathbf{p} \cdot \mathbf{M}_{ij}|^2 \delta(E_i - E_j + \hbar\omega), \quad (6)$$

where **p** is the light polarization vector. In this work, all the numerical results are performed by *Pybinding*.

## 3 Results and discussion

### 3.1 Energy spectrum

In Fig. 4, the energy levels of monolayer quantum dots of different shapes are calculated and plotted in the absence of an external field. Edge states emerge in the gap for the six QDs and the energy gap is around 1.8 eV. Comparing the eigenenergies of the six QDs above, we find that the distribution of states is similar in the valence band and the conduction band, while they vary in the band gap regime where edge states dominate. From the energy levels and DOS in the gap of PTQD, AsTQD and SQD, we could find that the energy of δ-P dominated states on ZZ edge occupy the higher energy part in the band gap regime and are completely detached from bulk states or other types of edge states, while the energy levels of δ-As dominated states on ZZ edge is about 0.7eV lower and are connected with the SZ and armchair edge states. |ψ|² on SZ and armchair edge states mainly dwelling on γ-P and γ-As atoms, and these kinds of edge states are connected with bulk states in the valence part. (2,1) RQD and (3,1) RQD also have δ-P atoms, but the energy levels of these states where the |ψ|² dwelling on the outmost δ-P atoms are much more centralized than normal ZZ edge states, as can be seen from the distinct peaks in DOS. This is because on these skewed edges, δ-P displayed far apart from each other and the couple of them is rather weak. So it is for the energy of γ-P and γ-As dominated states. We also find that the number of δ-P edge states is proportional to the number of δ-P atoms, for they are well detached from other kinds of states, while the δ-As, γ-P and γ-As edge states sometimes mix with each other.

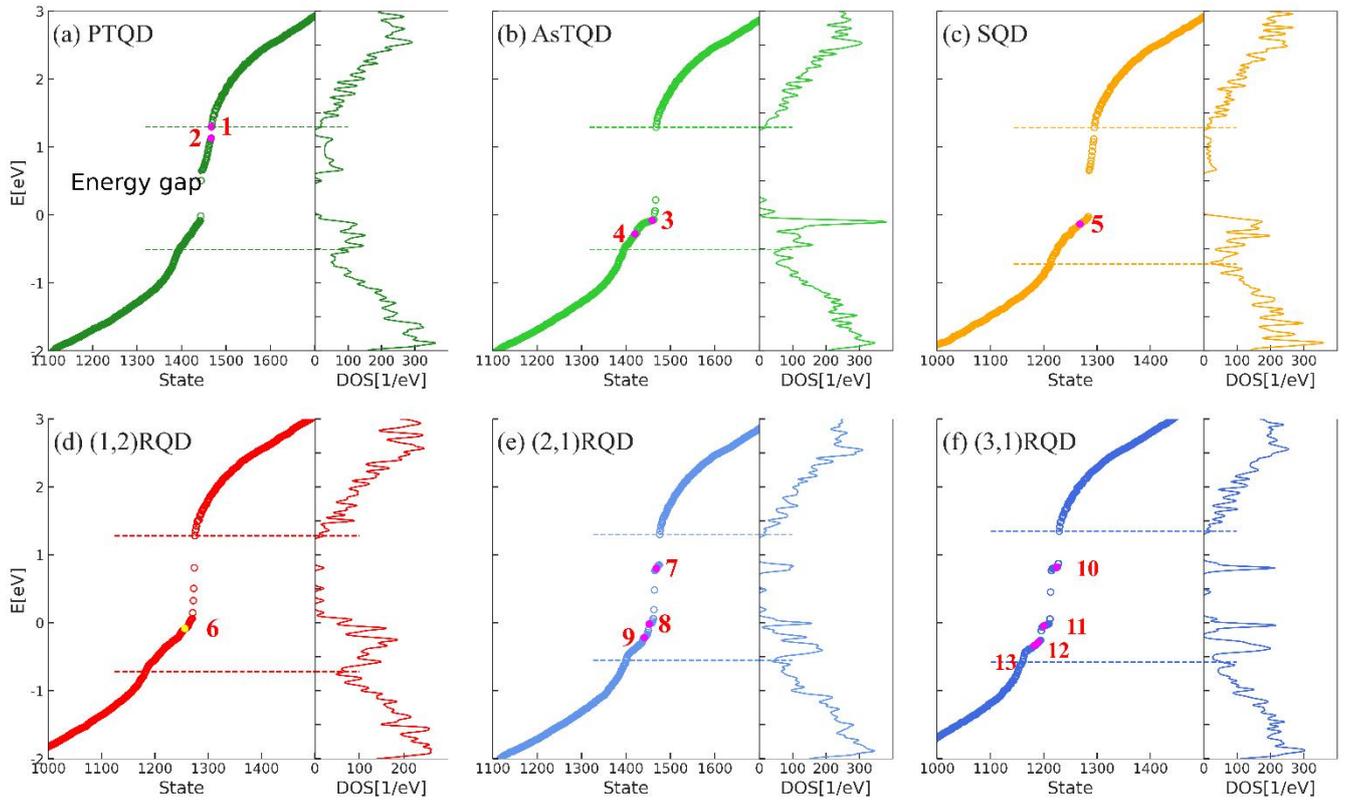

**Fig.4.** The energy levels of P/AsTQD, SQD, (1,2)/(2,1)/(3,1)RQD around the energy gap. The corresponding probability densities of the marked states are shown in Fig.5.

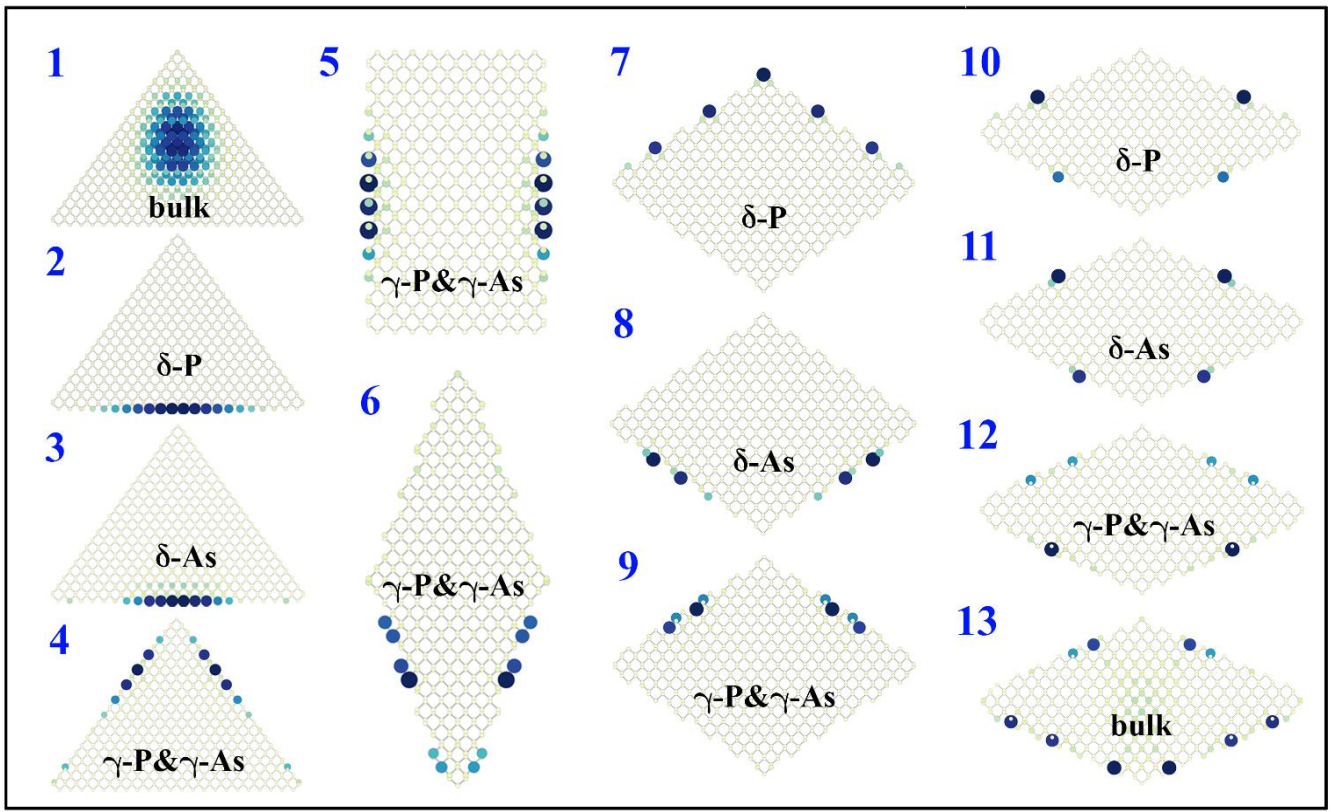

**Fig.5.** The corresponding probability densities of the states which are marked in Fig.4.

## 3.2 Electronic properties under an in-plane electric field

In this part, we investigate the electric field effect for SQD and (2,1) RQD, taking the center of SQD and (2,1) RQD as the zero of potential energy. The positive direction of the in-plane electric field and the corresponding force on electrons are shown in Fig.6(e). When the field is applied, the onsite energy of the atoms changes from $\varepsilon$ to $\varepsilon - e\mathbf{F} \cdot \mathbf{R}_{ij}$. As we can see from Fig.6(a), in the presence of $E_b$, the energy of $\delta$-P and $\delta$-As show a linear response to the increasing $E_b$ and they decrease/increase with the same ratio. This is because the electric field is perpendicular to these edges and their $|\psi|^2$ do not change with the increasing field, as a result, the onsite energy of $\delta$-P/$\delta$-As atoms decrease/increase linearly at the same ratio. However, when $E_b$ is applied to a (2,1) RQD, the edge states are no longer stable and the probability densities can be tuned by the field. As we can see from Fig.6.(b), the energy of upper and lower $\delta$-P dominated edge states decrease at different ratios and the 'edge bands' are no longer parallel. Since the skewed edges are not perpendicular to the field, under a small electric field, the probability densities of $\delta$-P edge states quickly reconstructed and the $|\psi|^2$ of the highest level moves along the positive direction of the electric field till they dwell on the two $\delta$-P atoms which have the highest potential energy while the $|\psi|^2$ of the lowest level move to the $\delta$-P atom in the corner where potential energy is the lowest. When we apply $E_a$ on SQD and (2,1) RQD, the energy of $\delta$-P edge states become decentralized and (2,1) RQD is obviously more sensitive to the changing $E_a$ for the $|\psi|^2$ quickly reconstructed at a rather small $E_a$ while the $|\psi|^2$ of SQD $\delta$-P edge states change gradually with the field. Their different sensitivity in the response of increasing $E_a$ is partly because the $\delta$-P atoms on skewed edges are separated apart. In the absence of an electric field, the energy levels are much more centralized than that of SQD, so a rather small $E_a$ can disturb it and lead to the redistribution of $|\psi|^2$. Under $E_a$, energy levels of $\gamma$-P and $\gamma$-As dominated edge states of SQD degenerate and split into two parallel bands while for (2,1) RQD they also split into two but in different directions, as is shown in the lower part of Fig.4(c)(d). Bulk states can also be pushed along the electric field.

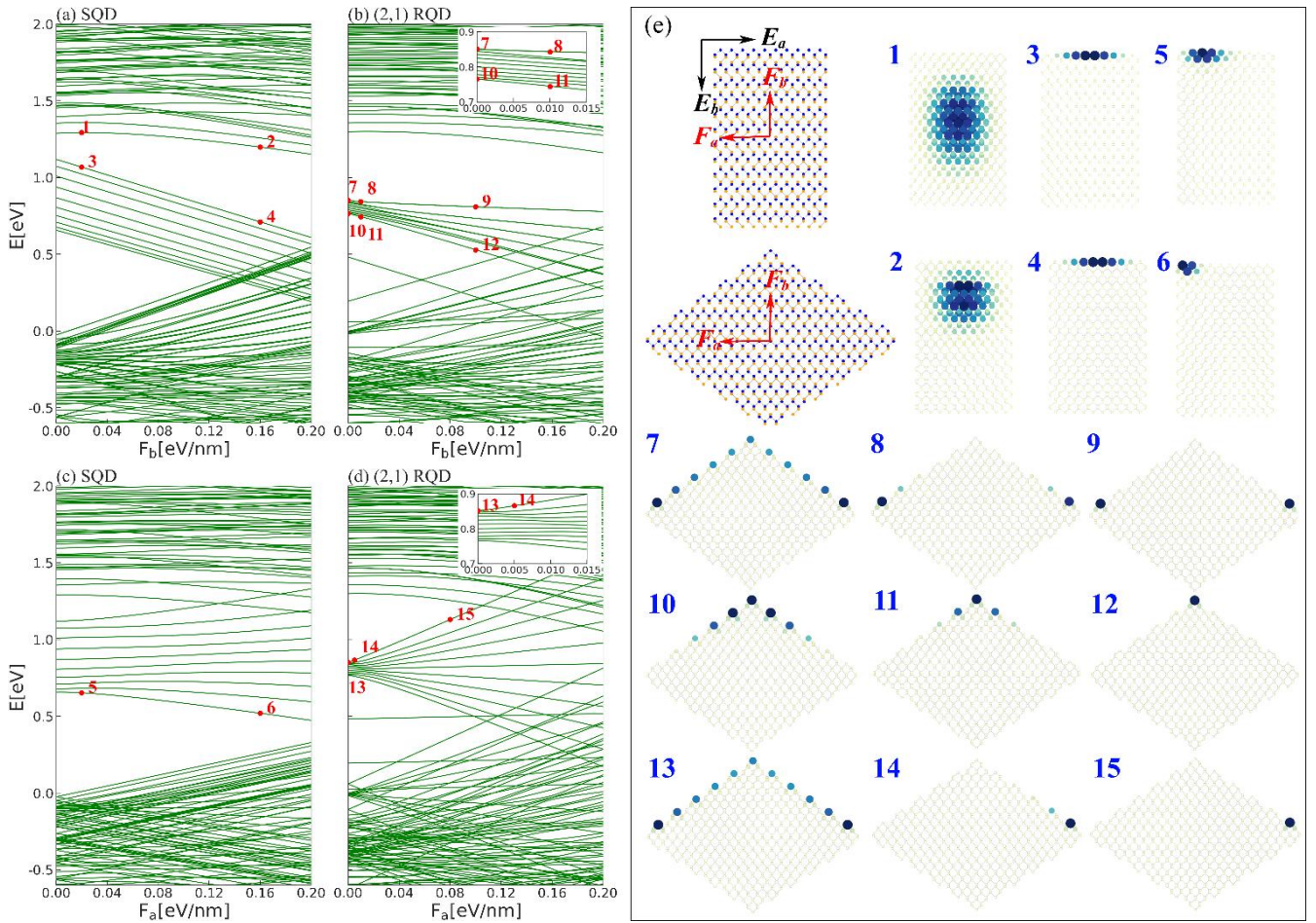

**Fig.6.** The energy spectra of SQD and (1,2)RQD as a function of in-plane *b*-direction and *a*-direction increasing electric field, respectively.

We can see from Fig.6(b) that the energy levels of $\delta$-P edge states degenerate at a relatively strong electric field, and Fig.6(e) shows that the probability densities can be tuned to specific atoms under the electric field. To further explore this phenomenon, we examine how the $\delta$-P edge states on skewed edges behave under a transverse electric field. Fig.7 and Fig.8 show the energy levels of δ-P dominated edge states and probability densities in (2,1)RQD, (2,1)RQR3 and (3,1)RQD. When the electric field is zero, (2,1) RQD has eleven δ-P edge states(which is equal to the number of δ-P atoms) and the corresponding $|\psi|^2$ are different and irregular. However, when $F_b$ increases to 0.2eV/nm, we can see clearly that the initial eleven levels degenerate to six, and the $|\psi|^2$ corresponding to these six levels dwell on two(or just one) δ-P atoms with the same potential. In this case we set the potential of the lower part of (2,1) RQD to zero(to avoid the energy intercross of δ-P and δ-As dominated edge states and just focus on δ-P dominated edge states), just adding $E_b$ to the upper part of it. When we apply $E_a$ to (2,1) RQD, we can see from Fig.7 (b) that the energy levels are still eleven, but they distribute nearly on a straight line, and the energy difference of adjacent levels is nearly equal to the potential difference of the adjacent δ-P atoms. The $|\psi|^2$ corresponding to these states are also quite interesting for they are just localized on a single atom with different potentials. The same is it for (2,1)RQR3 and (3,1)RQD, this implies that as long as a skewed outer or inner edge has δ-P atoms, such states may appear under the in-plane electric field. This does not happen in an SQD when we apply $E_a$ which is along its ZZ edge, for the distance between δ atoms on ZZ edges is much closer and the interaction between them is much stronger.

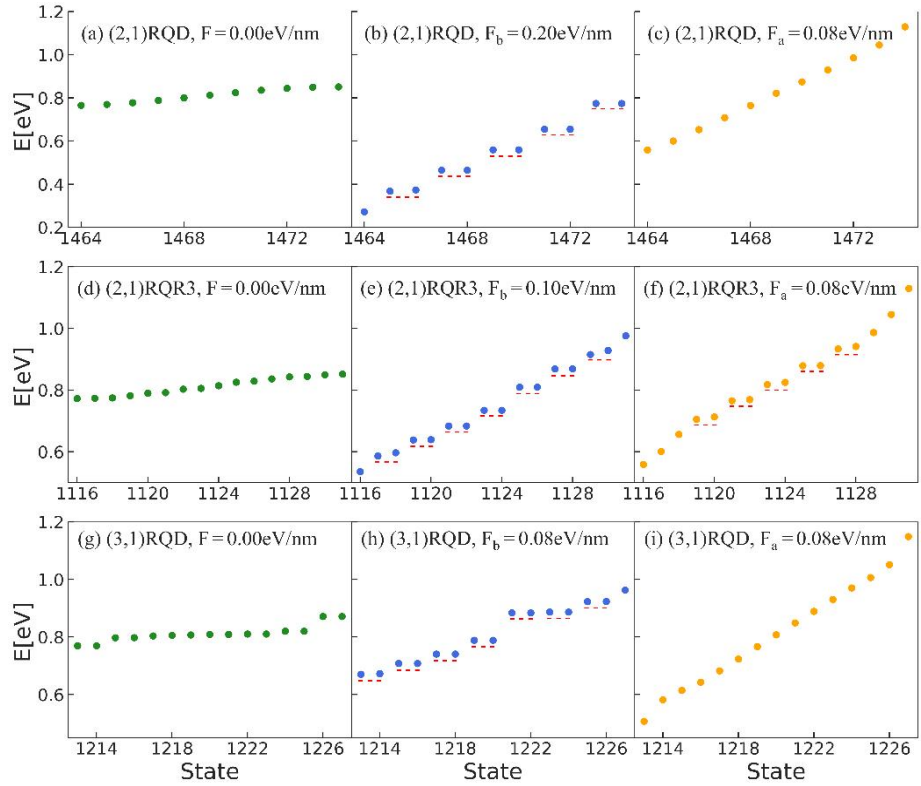

**Fig.7.** The energy levels of (2,1)RQD, (2,1)RQR3, and (3,1)RQD without external field and with transverse electric field along **a** and **b** directions.

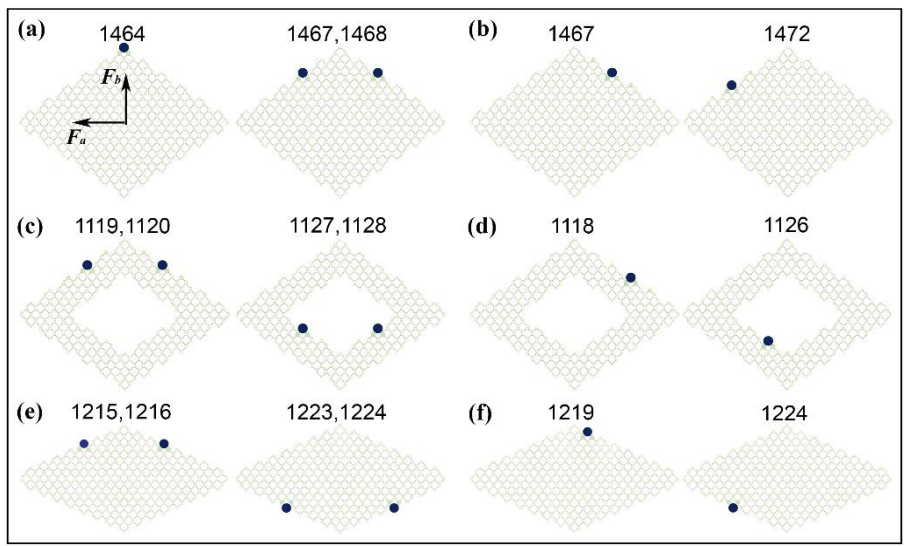

**Fig.8.** Probability densities of the corresponding energy levels in Fig.7, where (a)(b)(c)(d)(e)(f) corresponding to (b)(c)(e)(f)(h)(i)in Fig.7.

## 3.3 Electronic properties under the perpendicular magnetic field

In this part, we study the response of the edge states of (2,1) RQD/RQRs and (3,1) RQD under the perpendicular magnetic field. Fig.9 shows how the energy spectra are affected by the magnetic field. For (2,1) RQD, in the conduction regime, $|\psi|^2$ of the lowest bulk state first become more localized as the field increase due to the Lorenz effect but then split into two parts which may be caused by the Coulomb effect. In the energy region where $\delta$-P edge states dominate, the probability densities are nearly unaffected by the increasing perpendicular magnetic field and they form a quasi-flat energy band shown in Fig.9(b). However, in the same energy region of (3,1) RQD, they are no longer stable, as some of the energy levels first oscillate and then level off as a function of the magnetic field. This is because for a (2,1) RQD, the $\delta$-P atoms can only appear on two edges due to the chiral vector and edge configuration, while for a (3,1) RQD all the four edges have $\delta$-P atoms and they appear alternately. Therefore, the probability densities can roughly form a ring and they are weakly coupled, causing it to slightly oscillate and evolve due to the AB effect. As $\Phi$ increases to around $0.18\Phi_0$, the yellow-lined state in Fig.9(e) ceases to oscillate because due to anisotropy, the probability densities evolve to one side of the (3,1) RQD and no longer form a ring. (2,1) RQD and (3,1) RQD both have oscillating ring states in the valence region where $\gamma$ edge states dominate and the $|\psi|^2$ roughly form a ring. The oscillating period of (3,1) RQD is larger for it has a smaller area S. It can be seen clearly that the adjacent levels are anticrossing.

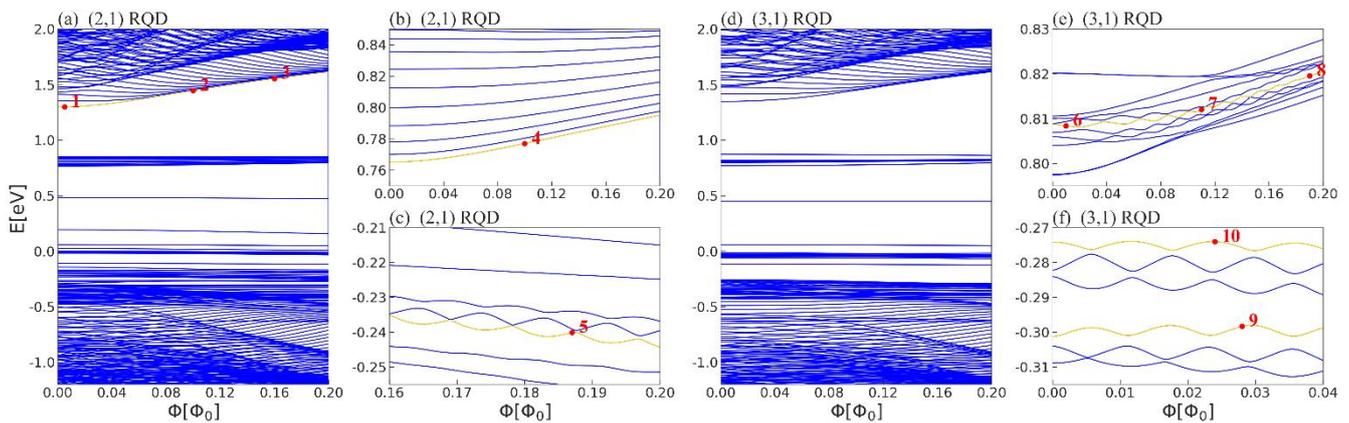

**Fig.9.** The energy spectra of (a)(2,1) RQD and (d)(3,1) RQD as a function of a perpendicular magnetic field. (b)(c)[(e)(f)] are the certain region of the spectrum of (2,1) RQD [(3,1) RQD] enlarged.

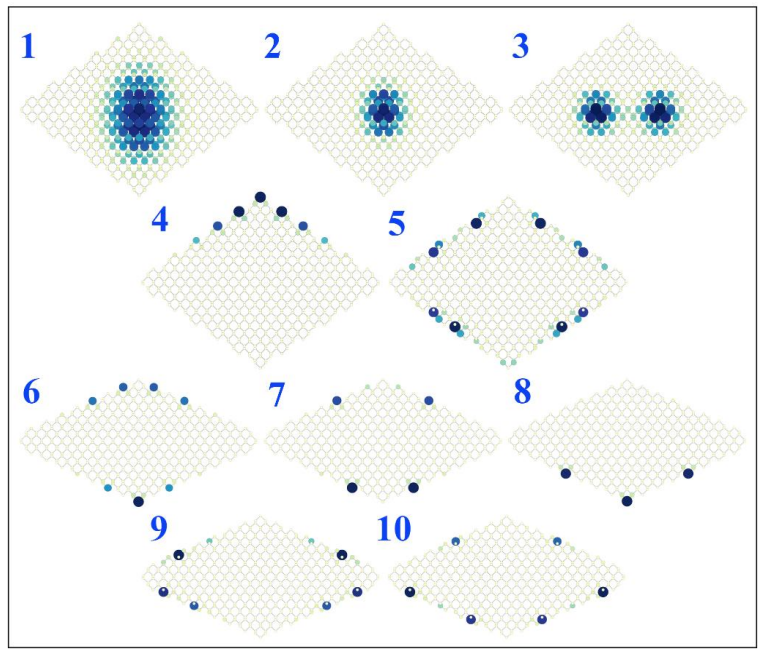

**Fig.10.** The corresponding probability densities of the states marked by red dots and numbers in Fig. 9.

We next calculate the energy spectra of (2,1)RQRs under an increasing perpendicular magnetic field. The basic information of the (2,1) RQRs is illustrated previously in Fig.3. Compared with (2,1) RQD, rich oscillating states emerge in the region where bulk states dominate. This is because the rings show enhanced confinement due to their inner edges and an inter-edge coupling, which strongly affects the energy spectrum and the wave functions. We can see from Fig.11(d)(e)(f) that the oscillating period decreases as the antidot size increases. Some energy levels first oscillate and then converge to several different Landau levels, while some levels are highly degenerated in the absence of a field but begin to oscillate and separate as the field increase.

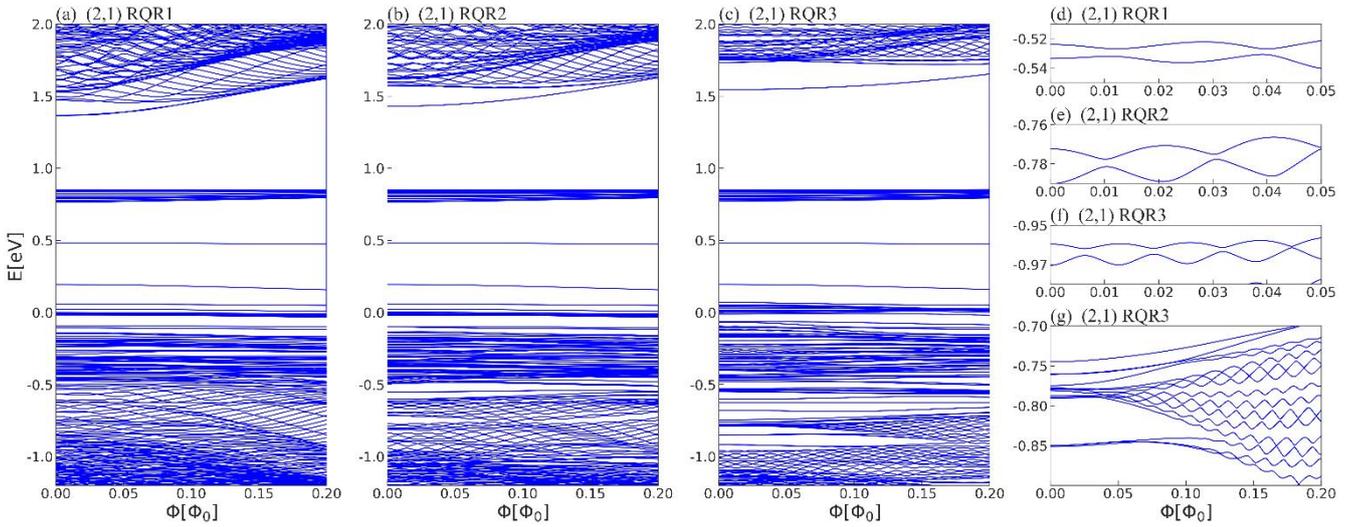

**Fig.11.** The energy spectra of (1,2) RQR of different antidot sizes as a function of the perpendicular magnetic field, respectively.

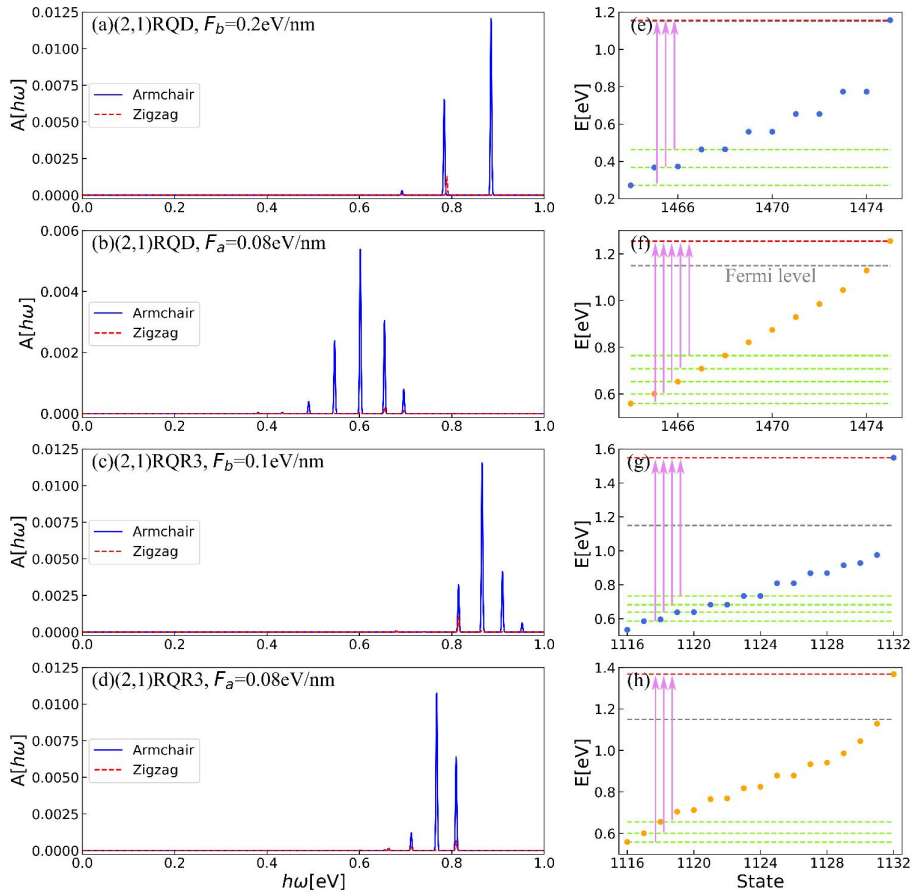

**Fig.12.** Optical absorption spectra of (2,1)RQD[(a)(b)], (2,1)RQR3[(c)(d)]. In (e)(f)(g)(h), the grey dashed line indicates the Fermi level, which is tuned to 1.15eV between the lowest bulk state and the highest δ-P dominated edge

state, the purple arrows indicate the transitions between the lowest bulk state (right dashed line) and the $\delta$-P dominated edge states(green dashed line), which are corresponding to the absorption peaks in the left panel. The broadening factor $\Gamma$ is set to 0.002eV.

## 3.4 Optical absorption

In this part, we examine the optical properties of (2,1)RQD, (2,1)RQR3 and (3,1)RQD. Fig.12 shows the optical absorption spectra of the QDs/QR in the presence of an in-plane electric field. Here we tune the Fermi energy inside the gap by a gate voltage and focus on the transition between $\delta$-P edge states and the lowest bulk state in the conductance region. Distinct and nearly evenly distributed absorption peaks appear in the spectra, which are corresponding to the transition between $\delta$-P edge states and the lowest bulk state in the conduction region, and the transition mainly happens in the $\delta$-P edge states with low energy. This may because $|\psi|^2$ of the lowest bulk state is pushed along the electric field and reached a place with lower potential energy, so it is closer to the $\delta$-P dominated edge states with low energy(as we have discussed before, $|\psi|^2$ of these states also dwell on the atoms where potential energies are lower). We can also find that the absorption mainly occurs when the light is polarized along the armchair direction due to the high anisotropy of monolayer b-AsP.

# 4 Summary

We have theoretically investigated the electronic and optical properties of b-AsP QDs with different edge types and (2,1) RQR with different antidot sizes in the presence of the perpendicular magnetic field as well as the in-plane electric field. Our result demonstrates that the eigenenergies and the evolution of edge states under the external field largely depend on the edge types and atom types. An in-plane electric field can tune the probability densities along the field direction, and for (2,1)RQD, the $|\psi|^2$ can be tuned to certain atoms by the in-plane electric field. Under an increasing perpendicular magnetic field, whether an energy level will oscillate depend on whether $|\psi|^2$ can distribute in a way that they roughly form a ring. For (2,1)RQRs, the oscillation period decreases as the antidot size increases. The light absorption spectra of b-AsP QD and QRs are polarization sensitive, and the optical spectra of edge-bulk transition exhibit evenly separated distinct peaks due to the effect of the in-plane electric field. Our work suggests a potential system to implement quantum state transfer utilizing two-dimensional quantum dots.